\documentclass[11pt,a4paper]{article}
\usepackage[margin=25mm]{geometry}
\usepackage[T1]{fontenc}
\usepackage{lmodern,amsmath,amssymb,amsthm,mathtools,booktabs,graphicx,microtype,enumitem,array}
\usepackage[dvipsnames]{xcolor}
\usepackage[colorlinks=true,linkcolor=MidnightBlue,citecolor=MidnightBlue,urlcolor=MidnightBlue]{hyperref}
\usepackage[font=small,labelfont=bf]{caption}
\usepackage{placeins,needspace,etoolbox}
\hypersetup{pdftitle={Minimum cores in lattice graphs: zonotope structure and quartic growth},pdfauthor={Navid Mehrdad},pdfsubject={Minimum strong defensive alliances in lattice graphs}}

\setlist{itemsep=2pt,topsep=4pt}
\newtheorem{theorem}{Theorem}[section]
\newtheorem{lemma}[theorem]{Lemma}
\newtheorem{proposition}[theorem]{Proposition}
\newtheorem{corollary}[theorem]{Corollary}
\theoremstyle{definition}
\newtheorem{definition}[theorem]{Definition}
\newtheorem{conjecture}[theorem]{Conjecture}
\theoremstyle{remark}\newtheorem{remark}[theorem]{Remark}
\BeforeBeginEnvironment{theorem}{\Needspace{6\baselineskip}}
\BeforeBeginEnvironment{lemma}{\Needspace{6\baselineskip}}
\BeforeBeginEnvironment{proposition}{\Needspace{6\baselineskip}}
\BeforeBeginEnvironment{corollary}{\Needspace{6\baselineskip}}
\pretocmd{\section}{\Needspace{6\baselineskip}}{}{}
\pretocmd{\subsection}{\Needspace{5\baselineskip}}{}{}
\newcommand{\Z}{\mathbb Z}\newcommand{\R}{\mathbb R}
\newcommand{\conv}{\operatorname{conv}}\newcommand{\core}{\operatorname{core}}

\newcommand{\norm}[1]{\left\lVert#1\right\rVert}

\title{Minimum cores in lattice graphs:\\zonotope structure and quartic growth}
\author{Navid Mehrdad}
\date{6 October 2026}
\begin{document}
\maketitle
\begin{abstract}
A core of a graph is a nonempty vertex set in which every member has at least as many neighbours inside the set as outside. We study the minimum size $c_{m,n}$ of a finite core in the graph on $\Z^m$ joining points at sup-norm distance at most $n$. The local majority condition imposes a global geometric constraint: the convex hull of every core has an explicit lattice zonotope $H_n^{(m)}$ as a Minkowski summand. In the plane, this zonotope has width asymptotic to $6\zeta(3)n^3/\pi^2$. We prove that all its lattice points form a core and construct a shell with $O(n^4)$ points. Together with a packing argument, this gives $c_{2,n}=\Theta(n^4)$ and coordinate widths $\Theta(n^3)$ for every minimum core. We determine $c_{2,1}=12$ and $c_{2,2}=96$ by analytic arguments and prove uniqueness up to translation in both cases. Explicit forced-set and deficit certificates determine further minima within the prescribed zonotope hull. We distinguish these restricted results from global optimality and state the remaining geometric conjectures. We also give a direct reduction from \textsc{Clique} showing that the core problem is NP-complete on general finite graphs, which settles a conjecture of Hassanpour in the terms in which it was posed.
\end{abstract}

\section{Introduction}
A group can sustain an action under a local majority rule only if each of its members has enough support within the group. This is a persistence question: it concerns the survival of an existing group, rather than the initiation of a cascade. Hassanpour~\cite{hassanpour} formulated it in terms of the \emph{minimal core} of a network. The same condition defines a \emph{strong defensive alliance} in graph theory~\cite{kristiansen}; on graphs without isolated vertices it is also the condition for a $1/2$-cohesive set in the terminology of Morris~\cite{morris}.

\begin{definition}
Let $G=(V,E)$ be a locally finite simple graph. A finite nonempty set $S\subseteq V$ is a \emph{core} if
\begin{equation}\label{eq:core}
\deg_S(v)\ge\deg_{V\setminus S}(v)\qquad(v\in S).
\end{equation}
When finite cores exist, $\core(G)$ is the smallest cardinality of one. We use \emph{minimum} for cardinality minimization, to distinguish it from inclusion-minimality.
\end{definition}

Our principal object is the radius-$n$ lattice graph
\[
G_{m,n}:\quad V=\Z^m,\qquad x\sim y\ \Longleftrightarrow\ 0<\norm{x-y}_\infty\le n,
\]
where $m,n\ge1$. Write $c_{m,n}=\core(G_{m,n})$ and $B_r(p)=p+[-r,r]^m$. Every vertex has even degree $(2n+1)^m-1$, so \eqref{eq:core} is equivalent to
\begin{equation}\label{eq:threshold}
|S\cap B_n(p)|\ge q_{m,n}:=\frac{(2n+1)^m+1}{2}\qquad(p\in S).
\end{equation}
Closed-ball counts include the centre. In planar arguments we abbreviate $q=q_{2,n}=2n^2+2n+1$.

The degree bound gives $c_{m,n}\ge q_{m,n}$, improving by one the bound $((2n+1)^m-1)/2$ in~\cite[Chapter 3]{hassanpour}. In one dimension it is exact: an interval of $n+1$ lattice points is a core. In two dimensions the true order is larger. The boundary of a core must accommodate every direction present in the neighbourhood ball; this forces a hull whose width grows cubically in $n$ and a core whose cardinality grows quartically.

For each primitive lattice vector $d$, put $k_d=\lfloor n/\norm d_\infty\rfloor$. Choosing one representative from each pair $\{d,-d\}$, define
\begin{equation}\label{eq:zonotope}
H_n^{(m)}=\sum_{\substack{d\in\Z^m\ \mathrm{primitive}\!\!\pmod\pm\\0<\norm d_\infty\le n}}[0,k_dd].
\end{equation}
The choice of representatives changes the zonotope only by a lattice translation. We write $H_n=H_n^{(2)}$ and $W_m(n)$ for its width in a coordinate direction; symmetry makes the choice of coordinate immaterial.

The main results are as follows.
\begin{enumerate}
\item \textbf{A required hull.} If $S$ is a core of $G_{m,n}$, then $\conv(S)=H_n^{(m)}+Q$ for a lattice polytope $Q$. Every hull vertex forces all lattice points of the hull within distance $n$ of that vertex (Section~\ref{sec:geometry}).
\item \textbf{Exact small-radius values.} The planar minima are $c_{2,1}=12$ and $c_{2,2}=96$, with unique minimizers up to translation. At radius two, any core with an enlarged hull has at least 98 points (Section~\ref{sec:exact}).
\item \textbf{Constructions and growth.} Every lattice point of $H_n$ is an integral sum of its generating segments. Hence $H_n\cap\Z^2$ is a core for every $n$, and
\begin{equation}\label{eq:intro-bounds}
\left(\frac{4\zeta(3)}{\pi^2}+o(1)\right)n^4
\le c_{2,n}\le
\left(\frac{48\zeta(3)}{\pi^2}+o(1)\right)n^4.
\end{equation}
Every minimum planar core has coordinate widths $\Theta(n^3)$ (Section~\ref{sec:growth}).
\item \textbf{Restricted exact values and certificates.} Within a translate of $H_n$, the minima for $n=1,\ldots,5$ are $12,96,388,1040,2532$. For $n=6$, explicit certificates give $4552\le c^{H}_{2,6}\le4560$; the numerical record reports the exact value 4560. In three dimensions at radius one, the minimum within $H_1^{(3)}$ is 392 (Section~\ref{sec:certificates}).
\end{enumerate}
Here $c^H_{m,n}$ denotes minimization restricted to cores contained in a translate of $H_n^{(m)}$. This distinction matters: a restricted optimum is not automatically a global one.

Hassanpour conjectured that computing the minimal core of a general graph is NP-hard~\cite[fn.~41]{hassanpour}. In the alliance vocabulary this was already known~\cite{jamieson}; Section~\ref{sec:complexity} gives a short direct reduction from \textsc{Clique} for precisely the condition~\eqref{eq:core}, so that the conjecture is settled in the terms in which it was posed. The main contribution of the paper is the lattice geometry and its consequences, not a priority claim for general-graph hardness.

\section{Complexity on general graphs}\label{sec:complexity}
On a finite graph, \textsc{Minimum Core} asks whether a nonempty core of size at most $K$ exists. A proposed vertex set is a polynomial-time certificate, so the problem is in NP.

\begin{theorem}[NP-completeness]\label{thm:np}
\textnormal{\textsc{Minimum Core}} is NP-complete. Consequently computing $\core(G)$ on general finite graphs is NP-hard.
\end{theorem}
\begin{proof}
Reduce from \textsc{Clique}. Handle trivial instances directly, and let $(J,k)$ have $2\le k\le|V(J)|$ and maximum degree $\Delta\ge1$. Set $K=k(\Delta+1)$. Keep $J$, attach $\Delta$ pendant vertices to every original vertex, and add a clique $Q$ on $2K+2$ vertices. Join each original vertex $v$ to
\[
b_v=\Delta+2(k-1)-\deg_J(v)
\]
distinct vertices of $Q$. These numbers are nonnegative and no larger than $|Q|$. Every original vertex now has degree $2\Delta+2(k-1)$, so membership in a core requires at least $\Delta+k-1$ internal neighbours.

A $k$-clique in $J$, together with all its pendants, is a core of size $K$. Conversely, let $S$ be a core of size at most $K$. No vertex of $Q$ can belong to $S$, since its degree is at least $2K+1$ and it would need at least $K+1$ neighbours in $S$. A pendant can belong to $S$ only with its original vertex.

Put $T=S\cap V(J)$, $t=|T|$, and let $e(T)$ count its induced edges. The set $T$ is nonempty. Each $v\in T$ has at least $k-1$ neighbours in $T$, so $t\ge k$, and needs at least $\Delta+k-1-\deg_T(v)$ of its pendants. Therefore
\[
|S|\ge t(\Delta+k)-2e(T),
\qquad 2e(T)\ge(t-k)\Delta+k(t-1).
\]
The bound $2e(T)\le t\Delta$ gives $t\le\Delta+1$. If $t=k$, the displayed inequality forces all $\binom{k}{2}$ edges, so $T$ is a clique. If $t>k$, combining it with $2e(T)\le t(t-1)$ gives $(t-k)(t-1-\Delta)\ge0$, hence $t=\Delta+1$. The same inequality then forces $2e(T)=t(t-1)$, so $T$ is a clique of size at least $k$.

The construction is polynomial in the input size. Together with membership in NP this proves the theorem.
\end{proof}

\begin{figure}[htbp]\centering
\includegraphics[width=.8\textwidth]{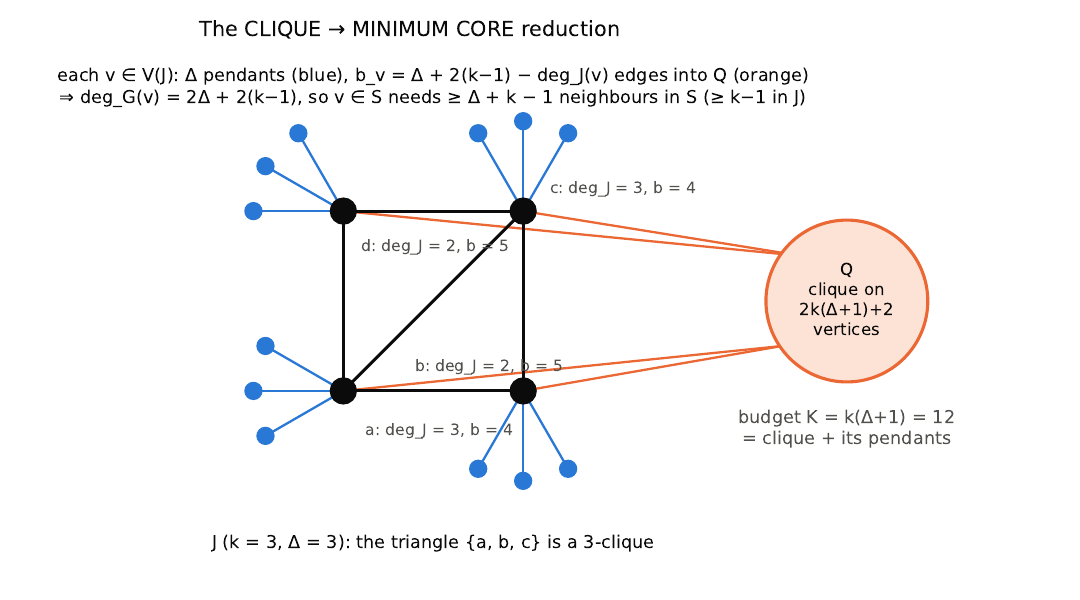}
\caption{The reduction of Theorem~\ref{thm:np} on a four-vertex instance ($k=3$, $\Delta=3$). Pendants (blue) inflate the degree of every original vertex so that membership in a small core requires at least $k-1$ neighbours among the original vertices; the heavy clique $Q$ (orange) absorbs the remaining degree without ever being affordable within the budget $K=k(\Delta+1)$.}\label{fig:reduction}
\end{figure}

\begin{remark}
(i) The construction was also checked mechanically on 263 random instances $(J,k)$ with $|V(J)|\le9$: an exact solver for the minimum core of the constructed graph agreed with $\omega(J)\ge k$ in every case. (ii) The ordinary defensive-alliance condition permits one more outside neighbour, $\deg_S(v)+1\ge\deg_{V\setminus S}(v)$~\cite{kristiansen}. On $G_{m,n}$ it coincides with the core condition because every degree is even; on arbitrary graphs the two must be kept distinct. The reduction adapts to the ordinary condition by replacing each pendant by a vertex adjacent to its original vertex and to two vertices of $Q$; it then has degree three and still forces its original vertex whenever $Q$ is excluded. (iii) In the parameterized sense the problem remains hard on structurally restricted inputs: it is W[1]-hard parameterized by treewidth~\cite{bliem}, while several other parameterizations are tractable~\cite{gaikwad}.
\end{remark}

\section{The geometry forced by a local majority}\label{sec:geometry}
A connected component of a core is itself a core: it contains every core-neighbour of each of its vertices. Therefore a minimum core is connected. We will use this observation for packing and finite-search bounds; the hull results in this section require no connectedness assumption.

\subsection{Extreme points and normal fans}
For a polytope $P$ and a vertex $v$, let $T_P(v)=\operatorname{cone}(P-v)$ be the tangent cone and let $N_P(v)$ be the open cone of linear functionals uniquely maximized at $v$. The cone $T_P(v)$ is pointed.

\begin{lemma}[Local half-ball forcing]\label{lem:local}
Suppose $p\in S$ uniquely maximizes a linear functional $\ell$ on $S\cap B_n(p)$, and $\ell\cdot u\ne0$ for every nonzero lattice vector $u\in B_n(0)$. Then every lattice point $p+u$ with $\ell\cdot u<0$ belongs to $S$.
\end{lemma}
\begin{proof}
The involution $u\mapsto-u$ partitions the nonzero ball vectors into two equal open half-balls. Only $p$ and the inward half-ball can belong to $S\cap B_n(p)$, giving exactly $q_{m,n}$ available positions. Equation~\eqref{eq:threshold} forces all of them.
\end{proof}

\begin{theorem}[Full corners and fan refinement]\label{thm:corner}
Let $P=\conv(S)$ for a core $S$. At every vertex $v$ of $P$:
\begin{enumerate}[label=(\roman*)]
\item each nonzero lattice ball vector $u$ belongs to exactly one of $T_P(v)$ and $-T_P(v)$;
\item $S\cap B_n(v)=B_n(v)\cap(v+T_P(v))\cap\Z^m$, and this set has exactly $q_{m,n}$ points.
\end{enumerate}
Consequently $P$ is full dimensional, and its normal fan refines the arrangement of the hyperplanes $d^\perp$ for primitive $d$ with $\norm d_\infty\le n$.
\end{theorem}
\begin{proof}
If $u\notin-T_P(v)$, some $\ell\in N_P(v)$ satisfies $\ell\cdot u<0$. Perturb $\ell$ within the open normal cone to avoid all finitely many ball hyperplanes. Lemma~\ref{lem:local} gives $v+u\in S$, so $u\in T_P(v)$. Thus the two cones contain at least one vector from every opposite pair; pointedness makes the choice exclusive. Applying the same argument to each $u\in T_P(v)$ forces the full corner and gives its cardinality.

In particular the cone contains one of $e_i,-e_i$ for every coordinate $i$, so it spans $\R^m$ and $P$ is full dimensional. On each open normal cone, the sign of $\ell\cdot d$ is fixed for every ball direction $d$. This is precisely the stated fan refinement.
\end{proof}

\begin{figure}[htbp]\centering
\includegraphics[width=\textwidth]{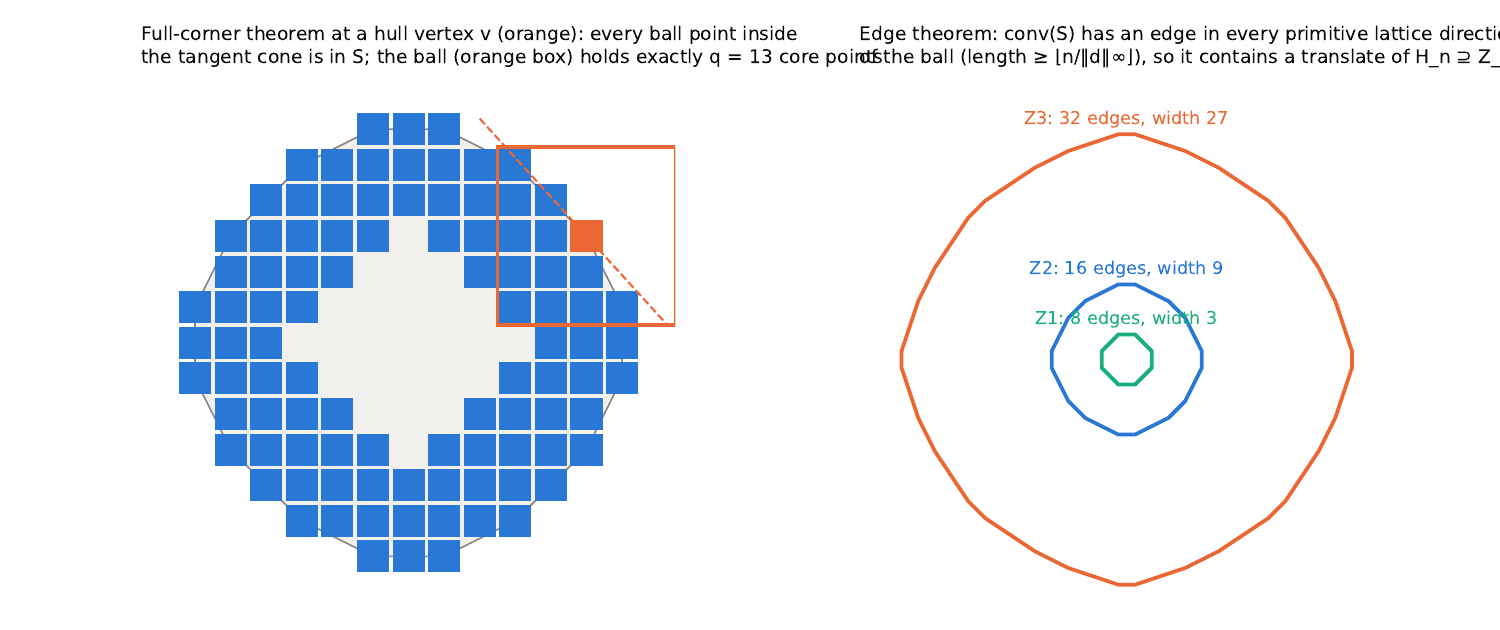}
\caption{Left: Theorem~\ref{thm:corner} on the radius-two minimiser. At the hull vertex $v$ (orange) every ball point inside the tangent cone belongs to the core, and the ball (orange box) contains exactly $q=13$ core points. Right: the zonotopes $Z_1\subset Z_2\subset Z_3$ generated by each primitive direction once; the hull of a core contains the larger zonotope $H_n$, in which direction $d$ is taken $k_d$ times (Theorem~\ref{thm:summand}).}\label{fig:structure}
\end{figure}

\begin{lemma}[Edge multiplicities]\label{lem:edges}
An edge of $P$ parallel to a primitive vector $d$ has lattice length at least $k_d$.
\end{lemma}
\begin{proof}
Write its endpoints as $v$ and $w=v+\lambda d$, where $\lambda$ is a positive integer. If $\lambda<k_d$, then $(\lambda+1)d\in B_n(0)\cap T_P(v)$. Theorem~\ref{thm:corner} forces $w+d\in S$, contradicting that $w$ is a vertex.
\end{proof}

\begin{theorem}[Zonotope summand]\label{thm:summand}
For every finite core of $G_{m,n}$,
\[
\conv(S)=H_n^{(m)}+Q
\]
for a lattice polytope $Q$, possibly a point or a lower-dimensional polytope.
\end{theorem}
\begin{proof}
We give a support-function proof that applies in every dimension. Put $P=\conv(S)$, $H=H_n^{(m)}$, and $f=h_P-h_H$, where $h_K(\ell)=\max_{x\in K}\ell\cdot x$. The normal fan of $H$ is the arrangement in Theorem~\ref{thm:corner}; hence $f$ is continuous and linear on every maximal cone of the normal fan of $P$.

Consider two adjacent maximal cones of that fan. Their corresponding vertices of $P$ differ by an edge vector $\lambda d$, oriented from the first cone to the second. If their common wall lies inside a maximal cone of the fan of $H$, the gradient of $h_H$ does not change, so the gradient jump of $f$ is $\lambda d$. Otherwise the wall lies in $d^\perp$ for a primitive ball direction, and the gradient jump of $h_H$ is $k_dd$. In this case the gradient jump of $f$ is $(\lambda-k_d)d$, a nonnegative multiple of the same oriented vector by Lemma~\ref{lem:edges}.

Thus the slope of $f$ does not decrease along a generic line crossing the walls. Lines meeting cones of codimension at least two follow by a limiting argument, so $f$ is convex. A convex, positively homogeneous, piecewise-linear function is the support function of the convex hull of its gradients. Those gradients are differences of lattice vertices of $P$ and $H$, so that convex hull is a lattice polytope $Q$. The identity $h_P=h_H+h_Q$ gives $P=H+Q$.
\end{proof}

For zonotopes and normal fans see~\cite{ziegler}. An alternative route to Theorem~\ref{thm:summand} is the Shephard--Weil--Schneider summand criterion~\cite[Thm.~3.2.11]{schneider}: a polytope $H$ is a Minkowski summand of $P$ if and only if $P(u)$ contains a translate of $H(u)$ whenever $H(u)$ is an edge, which is exactly what Theorem~\ref{thm:corner} and Lemma~\ref{lem:edges} provide.

\subsection{Forced sets and the boundary of the minimum hull}
For any lattice polytope $P$, define
\[
F_n(P)=P\cap\Z^m\cap\bigcup_{v\in\operatorname{vert}P}B_n(v).
\]
For a planar zonotope $H_n$, also define
\[
F_n^{\partial}(H_n)=H_n\cap\Z^2\cap\bigcup_{z\in\partial H_n\cap\Z^2}B_n(z).
\]
The latter union is over \emph{lattice} boundary points.

\begin{lemma}[Exact boundary counts]\label{lem:boundary}
Every $z\in\partial H_n\cap\Z^2$ satisfies $|B_n(z)\cap H_n\cap\Z^2|=q$.
\end{lemma}
\begin{proof}
Let $z$ lie on an edge with primitive direction $d$ and lattice length $k=k_d$, and let $\ell$ be its outer normal. Orient the zonotope generators so that their scalar products with $\ell$ are nonnegative, absorbing the resulting lattice translation. A point on the support edge uses the upper endpoint of every generator not parallel to $d$.

If $u$ is a ball vector with $\ell\cdot u<0$, write $u=-rd'$ with $d'$ primitive, $\ell\cdot d'>0$, and $1\le r\le k_{d'}$. Subtracting $rd'$ from its generator coefficient shows $z+u\in H_n$. The outward half-ball is outside. On the edge line, the ball contains $2k+1$ lattice points, of which exactly the $k+1$ points of the edge lie in $H_n$. The total is
\[
\frac{(2n+1)^2-(2k+1)}2+(k+1)=q.
\]
\end{proof}

\begin{proposition}[Forced sets]\label{prop:forced}
If $P=\conv(S)$, then $F_n(P)\subseteq S$. If $m=2$ and $P$ is a translate of $H_n$, then the corresponding translate of $F_n^{\partial}(H_n)$ is contained in $S$. Moreover, any core contained in a translate of $H_n^{(m)}$ has that entire translate as its convex hull.
\end{proposition}
\begin{proof}
Vertex forcing is Theorem~\ref{thm:corner}. Each edge of $H_n$ has sup-norm length $k_d\norm d_\infty\le n$, so every boundary lattice point is within a vertex ball and is forced. Lemma~\ref{lem:boundary} then forces its entire intersection ball.

For the final assertion, Theorem~\ref{thm:summand} puts a translate of $H_n^{(m)}$ inside the hull. A bounded full-dimensional convex body cannot properly contain a translate of itself: comparison of support functions would give $\ell\cdot t\le0$ for every $\ell$, hence $t=0$. The two inclusions therefore force equality.
\end{proof}

\section{The exact planar minima at radii one and two}\label{sec:exact}
We first obtain local counts near a coordinate extreme. Normalize the minimum occupied $x$-coordinate to zero, and let $(0,a)$ and $(0,b)$ be the lowest and highest occupied points in that column.

\begin{lemma}[Extreme-column forcing]\label{lem:extreme}
One has $b-a\ge n$, and $S$ contains
\[
\{(0,a+j):0\le j\le n\}\ \cup\
\{(i,a+j):1\le i\le n,\ -n\le j\le n\}.
\]
The reflected statement holds at $(0,b)$.
\end{lemma}
\begin{proof}
In the ball of $(0,a)$, negative columns and points below $a$ in column zero are unavailable. Exactly $(n+1)+n(2n+1)=q$ positions remain, so all are forced. In particular $(0,a+n)\in S$.
\end{proof}

\begin{lemma}[Two end-strip bounds]\label{lem:strips}
Every planar core contains at least $q+n^2=3n^2+2n+1$ points in its first $n+1$ columns and at least
\[
C_n=2q+n^2-1=5n^2+4n+1
\]
points in its first $n+2$ columns. The same bounds hold at the other three coordinate extremes. For $n=2$, the latter bound is 29, and it rises to 32 if the extreme hull edge has lattice length at least three.
\end{lemma}
\begin{proof}
For the first bound, column zero contains at least $n+1$ points. In each column $1,\ldots,n$, the endpoint-forced intervals $[a-n,a+n]$ and $[b-n,b+n]$ have a union of at least $3n+1$ lattice points because $b-a\ge n$. Thus the count is at least $(n+1)+n(3n+1)=q+n^2$.

For the second bound, the forced points $p=(1,a-n)$ and $r=(1,b+n)$ have disjoint radius-$n$ balls, each containing at least $q$ core points, all in columns $0,\ldots,n+1$. The additional forced rectangle
\[
\{0,\ldots,n\}\times\{a+1,\ldots,a+n-1\}
\]
lies strictly between the balls and has $n^2-1$ points. If $n=2$ and $b-a\ge3$, the three points in row $a+2$, columns $0,1,2$, are also forced and lie strictly between the balls.
\end{proof}

\begin{theorem}[Radius one]\label{thm:n1}
The minimum is $c_{2,1}=12$, and its unique minimizer up to lattice translation is $H_1\cap\Z^2$.
\end{theorem}
\begin{proof}
The summand theorem gives coordinate widths at least $W_2(1)=3$. The first two and last two columns are therefore disjoint, and each contains at least six points by Lemma~\ref{lem:strips}. Hence every core has at least 12 points. The set $H_1\cap\Z^2$ is the $4\times4$ lattice square with its four corners removed; it has 12 points and minimum closed-ball count five.

A 12-point core is connected. If its width exceeded three, the two end strips would already contain all 12 points and an empty intervening column would disconnect it in $G_{2,1}$. Thus both coordinate widths equal three. The summand theorem then makes $Q$ a point, and Proposition~\ref{prop:forced} forces all 12 points of the translated $H_1$.
\end{proof}

For radius two, normalize $H_2\subseteq[0,12]^2$. Its axis-normal edges have lattice length two, and it satisfies the four reflected corner inequalities represented by $x+y\ge4$. Write $P=H_2+Q$, where $Q$ has bounding rectangle $[0,u]\times[0,v]$. Let $M$ count the four axis-normal support faces of $Q$ that are nondegenerate segments. Let $R$ count the four labeled corners of its bounding rectangle absent from $Q$; retain all four labels when the rectangle is degenerate.

\begin{lemma}[Four strips]\label{lem:fourstrips}
Every radius-two core with hull $H_2+Q$ satisfies $|S|\ge92+3(M+R)$.
\end{lemma}
\begin{proof}
Take the first and last four columns and the first and last four rows, intersected with $S$. Each contains at least 29 points. When the corresponding support face of $Q$ is nondegenerate, the support edge of $P$ has length at least $2+1=3$, so its strip contains at least 32. Their cardinalities sum to at least $116+3M$.

Opposite strips are disjoint because both widths are at least 12. Their only intersections are four corner $4\times4$ boxes, with no triple intersections. At the lower left, put $\delta=\min_Q(x+y)$. If $(0,0)\in Q$, then $\delta=0$; otherwise $\delta\ge1$ by integrality. Thus $P\subseteq\{x+y\ge4+\delta\}$. In $\{0,1,2,3\}^2$ there are six points with $x+y\ge4$ and three with $x+y\ge5$. Reflections give total overlaps at most $24-3R$. Inclusion--exclusion yields the result.
\end{proof}

\begin{lemma}\label{lem:nonpoint}
If $Q$ is not a point, then $M+R\ge2$.
\end{lemma}
\begin{proof}
If $R\ge2$, this is immediate. If $R=1$, the bounding rectangle is nondegenerate and three of its corners belong to $Q$; an adjacent pair provides an axis-normal support segment. If $R=0$, convexity makes $Q$ its entire bounding rectangle, and any nonpoint rectangle, including an axis-parallel segment, has at least two nondegenerate axis-normal support faces.
\end{proof}

\begin{theorem}[Radius two]\label{thm:n2}
One has $c_{2,2}=96$. Every core whose hull is not a lattice translate of $H_2$ has at least 98 points. Every 96-point core is a lattice translate of $F_2^{\partial}(H_2)$.
\end{theorem}
\begin{proof}
A nonpoint summand $Q$ gives $|S|\ge98$ by Lemmas~\ref{lem:fourstrips} and~\ref{lem:nonpoint}. If $Q$ is a point, Proposition~\ref{prop:forced} forces $F_2^{\partial}(H_2)$. Its 96 points are specified in Table~\ref{tab:witness96}; direct row counts give at least 13 points in every radius-two closed ball. It is therefore a core, and equality makes the forced inclusion an equality.
\end{proof}

\begin{table}[htbp]\centering
\caption{The radius-two minimizer. Row $12-y$ repeats row $y$, and intervals include both endpoints.}\label{tab:witness96}
\begin{tabular}{@{}clrr@{}}\toprule
$y$&Occupied $x$-coordinates&Count&Minimum ball count\\\midrule
0&$[5,7]$&3&13\\
1&$[3,9]$&7&13\\
2&$[2,10]$&9&13\\
3&$[1,5]\cup[7,11]$&10&13\\
4&$[1,4]\cup[8,11]$&8&15\\
5&$[0,3]\cup[9,12]$&8&13\\
6&$[0,2]\cup[10,12]$&6&13\\\bottomrule
\end{tabular}
\end{table}
\begin{figure}[!htbp]\centering
\includegraphics[width=.62\textwidth]{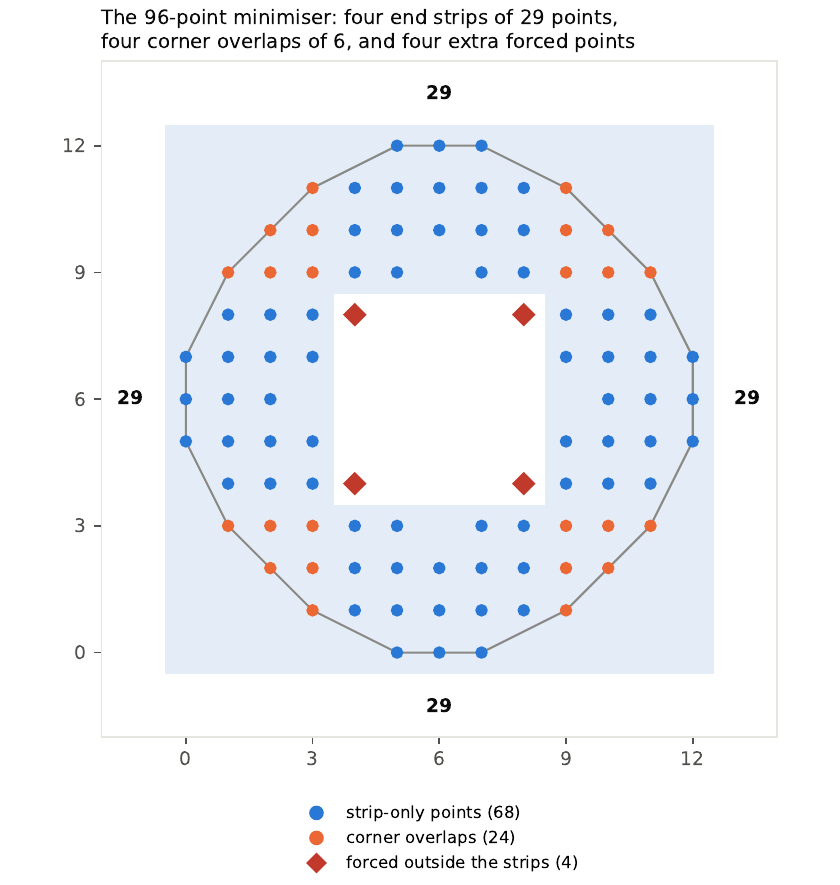}
\caption{The 96-point minimizer in the four-strip accounting of Lemma~\ref{lem:fourstrips}. Each shaded end strip contains 29 points; the four six-point corner overlaps (orange) leave 92 points in the union of the strips. The four diamonds lie outside the strips and are forced by the balls of the boundary midpoints $(2,2),(2,10),(10,2),(10,10)$. For an enlarged hull, longer extreme edges or smaller corner overlaps raise the bound to at least 98.}\label{fig:strips}
\end{figure}

\section{Finite constructions and growth}\label{sec:growth}
\subsection{Integral representations in the planar hull}
The next lemma allows us to build the lattice points of $H_n$ together with its generating segments. A lattice polytope here may be a point or a segment.

\begin{lemma}[Adding a unimodular parent sum]\label{lem:parents}
Let $a,b\in\Z^2$ satisfy $|\det(a,b)|=1$, let $d=a+b$, and let $K$ be a lattice polytope. For $P=K+[0,a]+[0,b]$,
\[
(P+[0,d])\cap\Z^2=(P\cap\Z^2)\cup\bigl((P\cap\Z^2)+d\bigr).
\]
If instead $P=K+[0,d]$ with $K$ any convex polytope and $d\ne0$, then $P+[0,d]=P\cup(P+d)$.
\end{lemma}
\begin{proof}
Use the lattice basis $a,b$, so that $d=(1,1)$ and $P=K+[0,1]^2$. Let $\ell(x,y)=x-y$, $\alpha=\min_K\ell$, and $\beta=\max_K\ell$. Both extrema are integers. A lattice point lies on a line $\ell=k$ with integer $k$.

For $\alpha\le k\le\beta$, the slice of $K$ is nonempty. Adding the diagonal of the unit square shows that the slice of $P$ is an interval of length at least one in the primitive $d$ coordinate. Such an interval $I$ satisfies $I+[0,1]=I\cup(I+1)$. The only remaining integer-level slices are $k=\alpha-1$ and $k=\beta+1$. Each is a support face of $K$ translated by a vertex of the square, so it has lattice endpoints; even for a singleton, its expansion introduces no lattice points beyond the original integer interval and its unit translate. This proves the first identity.

For the second, each nonempty slice parallel to $d$ is already an interval of length at least one in that coordinate, so the real-interval identity applies on every line.
\end{proof}

\begin{theorem}[Integral generator representation]\label{thm:integral}
Every $p\in H_n\cap\Z^2$ has a representation
\[
p=\sum_{d/\pm}c_dd,\qquad c_d\in\{0,1,\ldots,k_d\}.
\]
\end{theorem}
\begin{proof}
Choose representatives in the open upper half-plane together with $e_1$, and start with $[0,e_1]+[0,e_2]$. Every lattice point of this square is a sum of segment endpoints.

Every primitive $(r,s)$ with $r,s>0$ has nonnegative integer parents $a=(u,v)$, $b=(r-u,s-v)$ with $us-vr=1$. They are nonzero, primitive, lie in the same coordinate box as $(r,s)$, and have smaller coordinate sum. To see existence, if $r>1$ choose $u\in\{1,\ldots,r-1\}$ solving $us\equiv1\pmod r$ and put $v=(us-1)/r$; if $r=1$, take $(u,v)=(1,s-1)$. Reflection gives parents in the second quadrant. The possible parent $-e_1$ is available by reversing $[0,e_1]$ and translating the remaining summand by a lattice vector.

Add each primitive direction of sup-norm at most $n$ once, in increasing order of $|d_x|+d_y$. Its two parents are already present, and removing their segments leaves a lattice zonotope. The first identity in Lemma~\ref{lem:parents} preserves representation by segment endpoints. Finally add the remaining $k_d-1$ copies of each direction. The second identity preserves the property, and grouping equal directions gives the asserted coefficients.
\end{proof}

\begin{corollary}[The full planar hull is a core]\label{cor:fullhull}
For every $n\ge1$, $H_n\cap\Z^2$ is a core, and its minimum closed-ball count is exactly $q$.
\end{corollary}
\begin{proof}
For a representation in Theorem~\ref{thm:integral}, the points $p-jd$ for $1\le j\le c_d$ and $p+jd$ for $1\le j\le k_d-c_d$ stay in $H_n$. They supply $k_d$ distinct neighbours at distance at most $n$ in direction $d$. Distinct primitive directions modulo sign give distinct neighbours, and $\sum k_d=((2n+1)^2-1)/2=q-1$. Equality of the minimum follows from Lemma~\ref{lem:boundary}.
\end{proof}

The integral representation is a statement about decomposition into the \emph{specified generating segments}. It is stronger than the mere existence of real zonotope coordinates and should not be confused with the usual integer decomposition property for dilates of one polytope.

\subsection{Shells and existence in every dimension}
\begin{proposition}\label{prop:doubled}
The lattice points of $2H_n^{(m)}$ form a finite core for every $m,n\ge1$.
\end{proposition}
\begin{proof}
Write $p=\sum t_dd$ with $0\le t_d\le2k_d$. For each direction, if $t_d\ge k_d$, then $p-rd$ stays in the zonotope for $1\le r\le k_d$; otherwise $p+rd$ does. These are lattice neighbours of $p$, distinct across directions, and their total is $\sum k_d=q_{m,n}-1$.
\end{proof}

\begin{lemma}[A shell inherits the core condition]\label{lem:shell}
If $P\subset\R^m$ is a convex polytope and $P\cap\Z^m$ is a core, then
\[
T(P)=\{p\in P\cap\Z^m:B_{2n}(p)\not\subseteq P\}
\]
is a core.
\end{lemma}
\begin{proof}
Let $p\in T(P)$. If $B_n(p)\not\subseteq P$, one of its lattice corners $z$ is outside $P$, since a convex set containing every corner contains the whole cube. Every $y\in B_n(p)\cap P\cap\Z^m$ has $\norm{z-y}_\infty\le2n$, and hence also belongs to $T(P)$. All neighbours needed at $p$ survive.

Otherwise $B_n(p)\subseteq P$. Some facet $a\cdot x\le b$ has $b-a\cdot p<2n\norm a_1$. For every lattice ball vector $u$ with $a\cdot u\ge0$, the point $p+u$ lies in $P$ and satisfies the same strict inequality, so belongs to $T(P)$. The closed half-ball contains at least $q_{m,n}$ lattice points by central symmetry.
\end{proof}

\begin{theorem}[Explicit upper bounds]\label{thm:upper}
For all $m,n\ge1$,
\[
c_{m,n}\le4mn\bigl(2W_m(n)+1\bigr)^{m-1}.
\]
In the plane the stronger bound
\begin{equation}\label{eq:planarupper}
c_{2,n}\le |T(H_n)|\le8n\bigl(W_2(n)+1\bigr)
\end{equation}
holds.
\end{theorem}
\begin{proof}
Apply Lemma~\ref{lem:shell} to $2H_n^{(m)}$, or to $H_n$ in the plane. To count the shell, erode successively by the coordinate segments $[-2n,2n]e_i$. Their Minkowski sum is $[-2n,2n]^m$, so the final body consists of points whose radius-$2n$ cube stays inside the original body. Each intermediate body is convex. On each coordinate lattice line, an erosion removes at most $4n$ lattice points from an interval. There are at most $(2W_m(n)+1)^{m-1}$ relevant lines in each direction, or $W_2(n)+1$ in the undoubled planar construction.
\end{proof}

The full planar hull generally has order $n^6$ lattice points; the shell, rather than the full hull, provides the quartic cardinality bound.

\subsection{Widths, packing, and asymptotics}
\begin{proposition}[Zonotope widths]\label{prop:width}
For fixed $m\ge2$,
\begin{equation}\label{eq:widthasym}
W_m(n)\sim\frac{2^{m-2}\zeta(m+1)}{\zeta(m)}n^{m+1}.
\end{equation}
In the plane, with $\varphi$ Euler's totient and $\varphi(1)=1$,
\begin{equation}\label{eq:widthexact}
W_2(n)=3\sum_{r=1}^n r\varphi(r)\left\lfloor\frac nr\right\rfloor
\sim\frac{6\zeta(3)}{\pi^2}n^3.
\end{equation}
Also $|\partial H_n\cap\Z^2|=4n(n+1)$.
\end{proposition}
\begin{proof}
Let $X_m(t)$ be the sum of $|d_1|$ over primitive directions modulo sign with $\norm d_\infty\le t$. M\"obius inversion gives
\[
X_m(t)=\frac12\sum_{k\le t}\mu(k)k\,s_k(s_k+1)(2s_k+1)^{m-1},
\qquad s_k=\lfloor t/k\rfloor.
\]
Indeed, the unrestricted sum of $|v_1|$ in $[-s,s]^m\cap\Z^m$ is $s(s+1)(2s+1)^{m-1}$, and writing a vector as $kd$ introduces the factor $k$. Consequently
\[
X_m(t)=\frac{2^{m-2}}{\zeta(m)}t^{m+1}+o(t^{m+1});
\]
the floor errors are $O(t^2\log t)$ for $m=2$ and $O(t^m)$ for $m>2$. Since widths add under Minkowski addition,
\[
W_m(n)=\sum_{j=1}^n X_m(\lfloor n/j\rfloor).
\]
The bound $X_m(t)=O(t^{m+1})$ permits summation of the asymptotic using the convergent series $\sum j^{-(m+1)}$, proving \eqref{eq:widthasym}.

In two dimensions, the primitive directions on the sup-norm shell $r$ contribute $3r\varphi(r)$ to $X_2(r)-X_2(r-1)$ (the case $r=1$ is checked directly). This yields \eqref{eq:widthexact}. Finally, each direction contributes two edges of lattice length $k_d$, so the boundary count is $2\sum k_d=4n(n+1)$.
\end{proof}

\begin{lemma}[Connected packing]\label{lem:packing}
A connected core of $G_{m,n}$ of coordinate width $L$ satisfies
\[
|S|\ge q_{m,n}\left\lceil\frac{L+n}{3n}\right\rceil.
\]
\end{lemma}
\begin{proof}
Consecutive occupied levels of that coordinate differ by at most $n$. Starting at the smallest level, repeatedly choose the smallest occupied level at least $2n+1$ beyond the preceding choice. Consecutive chosen levels differ by at most $3n$, and the tail beyond the last choice is at most $2n$. For $t$ choices, $L\le3n(t-1)+2n=3nt-n$. Choose one core point at each selected level. Their radius-$n$ balls are pairwise disjoint and each contains at least $q_{m,n}$ core points.
\end{proof}

\begin{theorem}[Growth and finite search]\label{thm:growth}
For fixed $m\ge2$,
\[
c_{m,n}\ge q_{m,n}\left\lceil\frac{W_m(n)+n}{3n}\right\rceil,
\qquad \Omega(n^{2m})\le c_{m,n}\le O(n^{m^2}).
\]
In the plane, \eqref{eq:intro-bounds} holds, so $c_{2,n}=\Theta(n^4)$. Every minimum planar core has coordinate widths $\Theta(n^3)$. With
\[
U_n=8n(W_2(n)+1),\qquad E_n=3n\left\lfloor\frac{U_n}{q}\right\rfloor-n+1,
\]
a minimum planar core can be translated into $\{0,\ldots,E_n-1\}^2$. Thus exact minimization is a finite problem on $O(n^6)$ sites.
\end{theorem}
\begin{proof}
A minimum core is connected. Combine Theorem~\ref{thm:summand}, Lemma~\ref{lem:packing}, and Proposition~\ref{prop:width} for the lower bound. Theorem~\ref{thm:upper} gives the upper bounds. In the plane $q\sim2n^2$ and $W_2(n)\sim6\zeta(3)n^3/\pi^2$, giving the constants in \eqref{eq:intro-bounds}. Finally, Lemma~\ref{lem:packing} implies $L\le3n\lfloor |S|/q\rfloor-n$ for each coordinate; insert $|S|\le U_n$.
\end{proof}

The finite-window statement is not a polynomial-time algorithm: the number of binary configurations still grows exponentially in the number of sites.

\begin{figure}[htbp]\centering
\includegraphics[width=.72\textwidth]{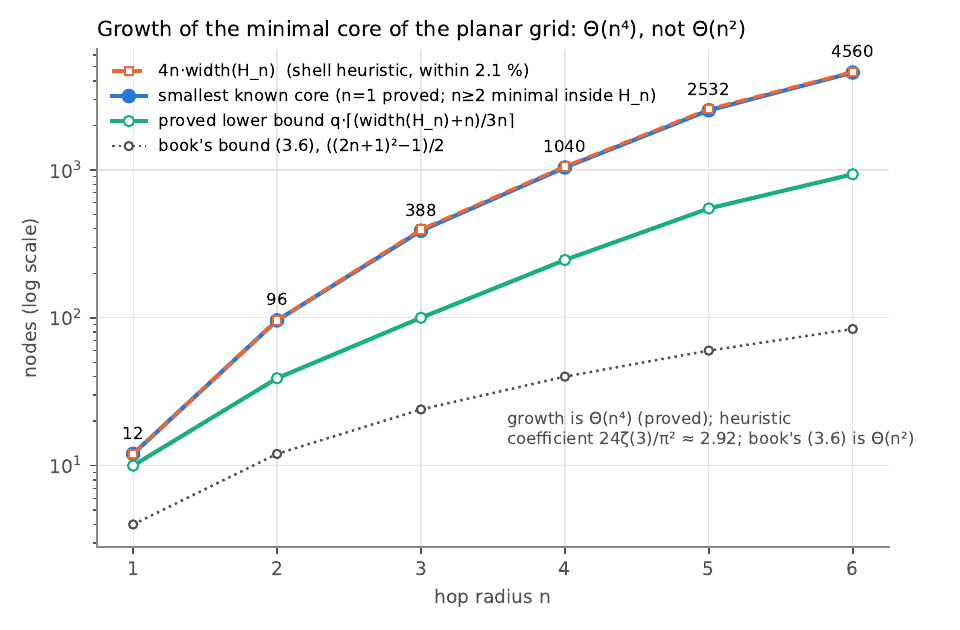}
\caption{The smallest exhibited cores (global minima for $n\le2$, hull-restricted minima for $n\ge3$) against the proved lower bound of Theorem~\ref{thm:growth}, the heuristic $4nW_2(n)$ of Section~\ref{sec:open}, and the bound of~\cite[eq.~3.6]{hassanpour}. Logarithmic scale.}\label{fig:growth}
\end{figure}

\begin{proposition}[A sharper planar width bound]\label{prop:sharperwidth}
If a connected planar core has coordinate width $L>5n+2$, then
\begin{equation}\label{eq:stripwidth}
|S|\ge2C_n+q\left\lceil\frac{L-5n-2}{3n}\right\rceil,
\qquad C_n=5n^2+4n+1.
\end{equation}
\end{proposition}
\begin{proof}
The two end strips occupy levels $0,\ldots,n+1$ and $L-n-1,\ldots,L$ and each contains at least $C_n$ points. Greedily choose occupied levels in $[2n+2,L-2n-2]$, separated by at least $2n+1$. The selected balls are disjoint from the strips and from one another. The first selected level is at most $3n+1$, each jump at most $3n$, and the last at least $L-5n-1$. Thus $t$ selections give $3nt\ge L-5n-2$.
\end{proof}

At radius two this gives $|S|\ge58+13\lceil(L-12)/6\rceil$: extents 26--31 require at least 97 points and extents 32--35 at least 110. In fact every extent at least 26 is excluded for connected cores of at most 96 points. The global proof of Theorem~\ref{thm:n2} is stronger and does not require these exclusions. At radius three, a core of at most 387 points contains a connected core with
\[
L\le17+9\left\lfloor\frac{387-116}{25}\right\rfloor=107,
\]
so a counterexample to the proposed global value 388 fits in a $108\times108$ window.

\section{Exact finite certificates and computational results}\label{sec:certificates}
Put $A_n=|F_n^{\partial}(H_n)|$. Proposition~\ref{prop:forced} gives $c^H_{2,n}\ge A_n$. All finite constructions and certificates asserted below can be checked with the ancillary standard-library Python program \texttt{verify\_minimum\_cores.py} (about four seconds). It uses exact integer geometry and counts; no optimizer is required. Universal results in the preceding sections follow from their proofs, not from finite tests.

\subsection{Forced sets and deficit certificates}
\begin{figure}[htbp]\centering
\includegraphics[width=\textwidth]{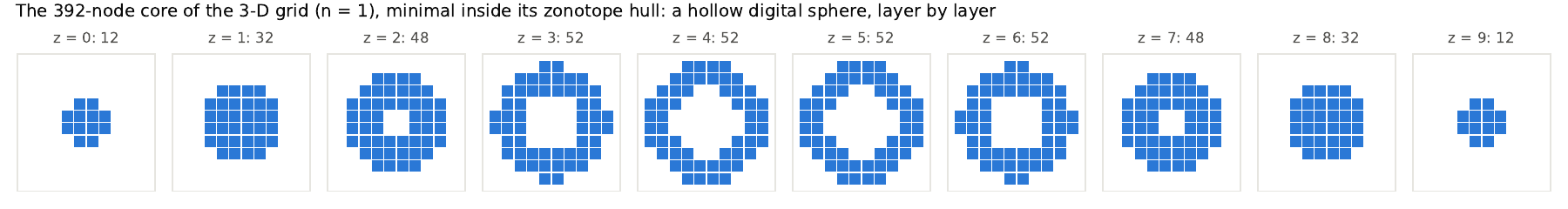}
\caption{The 392-point core of $G_{3,1}$, layer by layer: a hollow digital sphere in a $10^3$ box. It is the vertex-forced set of the zonotope $H_1^{(3)}$ and the unique minimum core inside that hull.}\label{fig:shell}
\end{figure}

For $n=1,2,3$, the forced sets have sizes $12,96,388$ and minimum ball counts $5,13,25$, respectively. Therefore they are the unique minimizers within their hulls. The corresponding three-dimensional forced set $F_1(H_1^{(3)})$ has 392 points and minimum ball count 14; its hull has 96 vertices and 480 lattice points (Figure~\ref{fig:shell}). This also gives uniqueness within its hull.

For larger radii the forced set may fail the core condition. For $p\in F=F_n^{\partial}(H_n)$ define its deficit
\[
\delta(p)=\max\{q-|F\cap B_n(p)|,0\}.
\]
If $S$ is a core inside $H_n$, then at least $\delta(p)$ points of $S\setminus F$ must lie in $B_n(p)$. Consequently, for any collection $C\subset F$ whose balls are pairwise disjoint,
\begin{equation}\label{eq:deficit}
|S|\ge |F|+\sum_{p\in C}\delta(p).
\end{equation}

Normalize $H_n$ to the box $[0,W]^2$, and let $\mathcal O_W(x,y)$ be the orbit under the eight symmetries of that square. Table~\ref{tab:deficits} specifies complete collections of centres; each displayed seed has an eight-point orbit, and all balls in the row are disjoint.
\begin{table}[htbp]\centering
\caption{Explicit disjoint-ball lower certificates. Each seed denotes its full square-symmetry orbit.}\label{tab:deficits}
\begin{tabular}{@{}rrrrlr@{}}\toprule
$n$&$W$&$A_n$&$q$&Seeds with deficits&Lower bound\\\midrule
4&66&1024&41&$(10,21):2$&1040\\
5&129&2452&61&$(12,46):2$, $(20,34):8$&2532\\
6&192&4392&85&$(19,62):3$, $(30,47):16$&4544\\\bottomrule
\end{tabular}
\end{table}

Appendix~\ref{app:witnesses} gives orbit representatives for additional points producing cores of sizes $1040$, $2532$, and $4560$. The first two match \eqref{eq:deficit}, proving $c^H_{2,4}=1040$ and $c^H_{2,5}=2532$. The radius-six case requires more than the forced-point constraints alone: the accompanying verifier contains an explicit set of 152 additional points satisfying all those constraints, but some added points see only six points in their radius-six balls. Thus the optimum of the forced-point-only integer relaxation is exactly $4392+152=4544$, while its witness is not a core.

\begin{figure}[!htbp]\centering
\includegraphics[width=\textwidth]{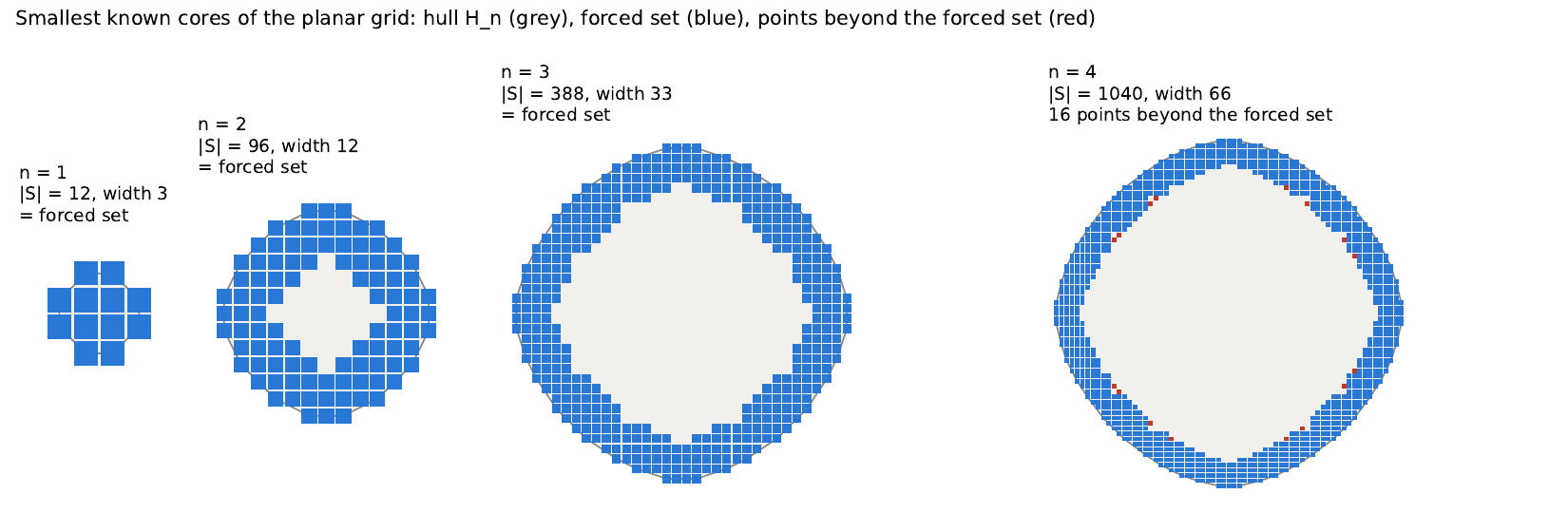}
\caption{Explicit planar cores at radii $1,2,3,4$, drawn to a common scale (hull $H_n$ in grey). Points of the forced set $F_n^{\partial}(H_n)$ are blue; the 16 additional points needed at radius four are red. The first two cores are global minima (Theorems~\ref{thm:n1} and~\ref{thm:n2}); the last two are proved minima within their displayed hulls.}\label{fig:cores}
\end{figure}

\subsection{An integer certificate for the radius-six lower bound}
The activated constraints at optional points give a stronger certificate. Centre $H_6$ at the origin, so its coordinate extrema are $\pm96$. Let $F=F_6^{\partial}(H_6)$ and, for every lattice point $p$ in the hull, set
\[
d(p)=85-|F\cap B_6(p)|.
\]
For each optional point $u\in(H_6\cap\Z^2)\setminus F$, let $x_u\in\{0,1\}$ record its inclusion. The following inequalities are necessary:
\begin{align}
\sum_{u\in B_6(p)\setminus F}x_u&\ge d(p)&& (p\in F),\label{eq:forcedconstraint}\\
\sum_{u\in B_6(p)\setminus F}x_u-d(p)x_p&\ge0&& (p\notin F),\label{eq:activated}\\
-x_u&\ge-1.\label{eq:uppervariable}
\end{align}
The sums run only over optional hull points. If an optional point is selected, \eqref{eq:activated} imposes its own core condition; if it is absent, the inequality is automatic.

\begin{proposition}[A 4552-point certificate]\label{prop:4552}
Every core contained in a translate of $H_6$ has at least 4552 points.
\end{proposition}
\begin{proof}
Use the full orbits under sign changes and coordinate interchange of the seven representatives in Table~\ref{tab:dual}. Multiply \eqref{eq:forcedconstraint} or \eqref{eq:activated}, as appropriate, by the listed integer. Also multiply \eqref{eq:uppervariable} by 1020 for the 16 points in the orbits of $(72,43)$ and $(71,43)$.

The sum has right-hand side 177376. Every optional-variable coefficient is between 0 and 1114; the positive coefficients lie between 34 and 1114. These integer coefficient bounds are verified directly by the supplied checker. Since $x_u\ge0$,
\[
1114\sum_u x_u\ge177376,
\qquad \sum_u x_u\ge\left\lceil\frac{177376}{1114}\right\rceil=160.
\]
Adding the 4392 forced points proves the claim.
\end{proof}

\begin{table}[htbp]\centering
\caption{The seven multiplier orbits in the radius-six certificate. Coordinates are centred, unlike those in Table~\ref{tab:deficits}.}\label{tab:dual}
\begin{tabular}{@{}crrr@{}}\toprule
Representative&In $F$?&$d(p)$&Multiplier\\\midrule
$(81,27)$&yes&3&94\\
$(75,37)$&yes&7&1020\\
$(70,43)$&no&17&60\\
$(69,43)$&no&30&34\\
$(66,49)$&yes&16&1020\\
$(59,57)$&yes&7&47\\
$(70,57)$&yes&3&47\\\bottomrule
\end{tabular}
\end{table}

\subsection{What the numerical record establishes}
For a finite window or prescribed hull $\Omega\subset\Z^m$, the exact binary model is
\begin{equation}\label{eq:ip}
\min\sum_{v\in\Omega}x_v,
\quad \sum_{u\in\Omega\cap B_n(v)}x_u\ge q_{m,n}x_v\ (v\in\Omega),
\quad \sum_{v\in\Omega}x_v\ge1,
\quad x_v\in\{0,1\}.
\end{equation}
Absent points outside $\Omega$ contribute zero. A feasible solution provides an upper bound. Global optimality requires either a proof for all possible hulls or a justified finite window containing every smaller connected core.

Table~\ref{tab:status} distinguishes explicit certificates included here from solver results recorded in the computational study. The latter used OR-Tools CP-SAT with eight workers; for the radius-six hull, a separate HiGHS mixed-integer solve also reported optimum 4560. The model after forcing $F$ has 22801 optional variables; both solvers reported that 168 additional points are necessary. The included certificate proves 160. The solver logs are kept with the computational record (Section~\ref{sec:data}) rather than in the verification package, so the sharper value is recorded separately from the independently checkable bounds.

\begin{table}[htbp]\centering\small
\caption{Status of the smallest exhibited cores. ``Reported'' denotes the solver record, beyond the explicit certificates supplied with this paper.}\label{tab:status}
\begin{tabular}{@{}ccrp{39mm}p{48mm}@{}}\toprule
$m$&$n$&Witness&Within $H_n^{(m)}$&Global status\\\midrule
1&any&$n+1$&Exact&Exact\\
2&1&12&Exact, unique&Exact, unique; Theorem~\ref{thm:n1}\\
2&2&96&Exact, unique&Exact, unique; Theorem~\ref{thm:n2}\\
2&3&388&Exact, unique&Open; reported optimal for extent $\le38$\\
2&4&1040&Exact&Open; reported optimal for extent $\le72$\\
2&5&2532&Exact&Open\\
2&6&4560&$[4552,4560]$ certified; reported exact 4560&Open\\
3&1&392&Exact, unique&Open; reported optimal for extent $\le12$\\\bottomrule
\end{tabular}
\end{table}

The former radius-two searches up to extent 25 are now corroboration only. Theorem~\ref{thm:n2} proves the global result without them, and Proposition~\ref{prop:sharperwidth} independently closes every extent at least 26 for a smaller connected candidate. At radius three, the reported extent-38 result leaves a genuine gap to the proved extent ceiling 108.

\section{Larger hulls and open problems}\label{sec:open}
The summand theorem reduces the global problem to understanding how a nontrivial lattice summand changes the minimum feasible core. At radius two, the four-strip argument answers this question without requiring a monotonicity theorem for forced sets. At larger radii, the following boundary accounting remains useful.

\subsection{Exact slack on a hull edge}
\begin{lemma}[Boundary slack]\label{lem:slack}
Let $P=H_n+Q$, and let a lattice point $z$ lie $j$ primitive steps from one endpoint of an edge of $P$ with direction $d$ and lattice length $\lambda$. Put $k=\lfloor n/\norm d_\infty\rfloor$, allowing $k=0$. Then
\[
|B_n(z)\cap P\cap\Z^2|=q+s_z,
\qquad s_z=\min(j,k)+\min(\lambda-j,k)-k.
\]
\end{lemma}
\begin{proof}
Let $\ell$ be the outward normal. Support faces add, so $z=h+q'$ with $h\in H_n(\ell)$ and $q'\in Q(\ell)$. The generator argument in Lemma~\ref{lem:boundary} shows that $h+u\in H_n$ for every inward ball vector $u$, hence $z+u\in P$. Outward vectors leave $P$. On the edge line, exactly $1+\min(j,k)+\min(\lambda-j,k)$ points remain. Counting the off-line half-ball proves the formula. The summand theorem gives $\lambda\ge k$, so $s_z\ge0$.
\end{proof}

Let $Z_n(P)=\partial P\cap F_n(P)$ be the boundary lattice points forced by hull vertices, and define
\begin{equation}\label{eq:tube}
\operatorname{LB}_n(P)=
\left|\bigcup_{z\in Z_n(P)}\bigl(B_n(z)\cap P\cap\Z^2\bigr)\right|
-\sum_{z\in Z_n(P)}s_z.
\end{equation}
Every core with hull $P$ has size at least $\operatorname{LB}_n(P)$: each selected ball may omit at most $s_z$ of its available points, and summing these omission allowances bounds the number absent from the union. For $P=H_n$, all slack vanishes and $\operatorname{LB}_n(H_n)=A_n$.

For an integer coordinate-segment enlargement, the bound has a closed form:
\begin{proposition}[Coordinate segments]\label{prop:segment}
For $t\in\Z_{\ge0}$ and $P_t=H_n+[0,t]e_1$,
\[
\operatorname{LB}_n(P_t)=A_n+2(n+1)\min(t,3n+1)-2n\min(t,n+1).
\]
The same holds with $e_2$. In particular, every $t\ge1$ gives $\operatorname{LB}_n(P_t)\ge A_n+2$.
\end{proposition}
\begin{proof}
The row-counting argument and slack sum are given in Appendix~\ref{app:segments}.
\end{proof}

\begin{conjecture}[A tube inequality]\label{conj:tube}
For every $n\ge1$ and every nonpoint lattice polytope $Q\subset\R^2$,
\[
\operatorname{LB}_n(H_n+Q)\ge A_n+2.
\]
\end{conjecture}
If true, this would prove $c_{2,3}=388$ and uniqueness of its minimizer, because $A_3=388$ and its forced set is a core. For $n\ge4$, the forced set alone is insufficient: deficits on enlarged hulls must also be controlled. In particular, the conjecture would not by itself establish the proposed global values 1040, 2532, and 4560.

Monotonicity under Minkowski addition is not supplied by the definition of a forced set. Different vertex neighbourhoods move by different translates, which may alter their overlaps. Even a result for every segment summand would not cover arbitrary lattice polygons, since a triangle is not a sum of segments. The radius-two theorem proves a lower bound for \emph{cores}; it does not establish the stronger geometric claim $\operatorname{LB}_2(H_2+Q)\ge98$ for all $Q$. Two pieces of evidence for the conjecture: for $n=2$ and every nonpoint $Q$ for which $H_2+Q$ is a zonotope with edge directions among the ball directions, $\operatorname{LB}_2\ge98$ (a saturation argument shows that once the two halves of the hull along a direction $d$ are $2n+1$ apart, lengthening the $d$-edges cannot decrease $\operatorname{LB}_2$, which reduces the statement to an exhaustive evaluation of $977{,}033$ configurations, over which $\operatorname{LB}_2$ is moreover coordinatewise non-decreasing); and for $n=3$, $\operatorname{LB}_3(H_3+Q)\ge390$ on $1243$ sampled enlargements. A stronger ``elongation'' statement, that each unit of extra width costs at least $2(n+1)$ points, is false: for $Q=[0,4]e_1$ at $n=2$ there is a $118$-point core with hull $H_2+Q$, against $96+24$.

\subsection{The remaining questions}
The most immediate global problem is the radius-three minimum. It has a complete hull-restricted certificate and a finite search ceiling, but neither the current tube argument nor the available solver record settles all larger hulls.

A second concrete problem is to replace the reported radius-six optimum by a short certificate for the last eight points beyond Proposition~\ref{prop:4552}. The 4544-point relaxation witness shows why constraints at newly selected points must enter such an argument.

For general $n$, define the thinner ring
\[
R_n^{\mathrm{thin}}=\{p\in H_n\cap\Z^2:B_{n+1}(p)\not\subseteq H_n\}.
\]
Exact checks for $1\le n\le10$ give cores of sizes
\[
12,108,464,1220,2952,5180,10016,15876,24920,35684.
\]
It remains open whether this ring is a core for every radius. Theorem~\ref{thm:upper} instead uses the proved thickness $2n$.

\begin{conjecture}[A minimum hull suffices]\label{conj:hull}
For every $n\ge3$, some minimum core of $G_{2,n}$ has hull a lattice translate of $H_n$.
\end{conjecture}
Even assuming this conjecture, the general hull-restricted minimum has no known closed formula. The six displayed examples lie near $4nW_2(n)$, whose asymptotic coefficient is $24\zeta(3)/\pi^2\approx2.923$. This is a heuristic suggested by the finite data; the rigorous coefficient bounds remain \eqref{eq:intro-bounds}, approximately $0.4872$ and $5.8461$. No limit for $c_{2,n}/n^4$ is proved here.

In higher dimensions, the gap is larger: at $m=3$ the present bounds are $\Omega(n^6)$ and $O(n^9)$. It is also open here whether the undoubled zonotope construction has the required integral representation in those dimensions.

\section{Discussion}
A local majority condition can impose a large geometric cost. At a hull vertex, exactly half of the available displacement directions point inward, so all corresponding neighbours must be present. These local requirements force every neighbourhood direction into the normal-fan structure, impose minimum edge multiplicities, and produce a mandatory zonotope summand. The cubic width and quartic core size in the plane follow from this geometry.

The exact values 12 and 96 illustrate two ways to use the structure: opposite end strips suffice at radius one, while radius two requires accounting for all four strips and their corner overlaps. Beyond radius two, the distinction between a forced set, a feasible core, and a globally minimum core becomes essential. The explicit deficit and activation certificates make part of that distinction quantitative.

For the collective-action interpretation, the proved conclusion is asymptotic: the least size of a sustainable group in this radius-$n$ planar family grows on the order of $n^4$. It is not a theorem of stepwise monotonicity in $n$, nor a claim that adding edges always enlarges a minimum core. On general graphs it can shrink: $K_{1,3}$ has minimum core size three, whereas adding an edge between two leaves creates a two-vertex core.

\section{Data and code}\label{sec:data}
The verification program accompanies this article as an ancillary file. The full computational record (solver scripts, solution files, logs of the CP-SAT and HiGHS runs, and the figure scripts) is available from the author.

\section*{Acknowledgements}
Several arguments in this paper were suggested in reports on earlier drafts, in particular the doubled-zonotope construction and shell lemma, the deficit certificates, the coordinate-segment formula, the end-strip lemmas, the four-strip argument of Theorem~\ref{thm:n2}, the integral-representation theorem, and the radius-six dual certificate; each has been verified independently here. Computations used OR-Tools CP-SAT 9.15 and HiGHS.

\appendix
\section{Explicit witnesses and verification}\label{app:witnesses}
This appendix specifies the finite data used by the restricted exact results. The accompanying program checks all stated ball counts, forced-set sizes, disjointness assertions, and dual coefficients using integer arithmetic. It does not rerun the external optimization searches or certify their logs.

\subsection{Constructing the hull and the forced set}
In the plane, list all primitive vectors $d$ with $0<\norm d_\infty\le n$ in counterclockwise order, including both signs, and use the edge vectors $k_dd$. Their sum is zero. Starting at the origin and cumulatively adding these edges constructs the hull up to translation; shift its coordinate minima to zero. For each integer row, intersect the exact supporting half-planes to obtain the lattice interval of the hull. Traverse each edge in primitive steps to list its boundary lattice points. The set $F_n^{\partial}(H_n)$ consists of those hull points within radius $n$ of this boundary list.

For reference, the planar forced-set sizes for $n=1,\ldots,10$ are
\[
12,96,388,1024,2452,4392,8412,13464,21172,30592.
\]
For $n=4,5,6$ their minimum closed-ball counts are $39,53,69$, below the thresholds $41,61,85$.

In three dimensions at radius one, take one vector from each opposite pair in $\{-1,0,1\}^3\setminus\{0\}$. Facet normals of the resulting zonotope are obtained from cross products of pairs of nonparallel generators. For each oriented normal $a$, its support value is $\sum_d\max(0,a\cdot d)$. These integer inequalities enumerate the hull exactly. A point is a vertex when its active facet normals span $\R^3$; this identifies the 96 vertices. Enumerating the 480 hull lattice points and their vertex balls gives the 392-point forced set and checks its threshold 14. No floating-point hull routine is needed.

\subsection{Additional points for radii four through six}
For each row below, add to $F_n^{\partial}(H_n)$ the full orbit $\mathcal O_W(p)$ of every displayed representative, using the uncentred box $[0,W]^2$. The orbits are disjoint and each has eight points.
\begin{itemize}
\item $n=4$, $W=66$: $(11,18),(11,19)$.
\item $n=5$, $W=129$:
\[
\begin{gathered}
(13,43),(13,44),(16,39),(18,35),(18,36),\\
(19,35),(21,30),(21,31),(22,30),(22,31).
\end{gathered}
\]
\item $n=6$, $W=192$:
\[
\begin{gathered}
(17,63),(17,64),(18,63),(22,57),(24,53),(24,54),(25,53),\\
(27,48),(27,49),(27,50),(27,51),(28,48),(28,49),(29,48),\\
(31,42),(31,44),(31,45),(32,42),(32,43),(32,44),(33,43).
\end{gathered}
\]
\end{itemize}
The resulting sets have $1024+16=1040$, $2452+80=2532$, and $4392+168=4560$ points. Every occupied point meets its corresponding core threshold. The 152-point witness for the forced-point-only relaxation is supplied as the literal list \texttt{BARRIER\_EXTRAS} in the verifier; its small ball count at some added points is the reason it is not a core.

\subsection{Scope of the checks}
Running \texttt{python3 verify\_minimum\_cores.py} verifies, for $n\le6$, the lattice point counts, widths and boundary counts of $H_n$, the exact boundary ball counts, the core property of the full hull and of the thin ring, the integral generator representations, the forced sets with their minimum ball counts, the witness cores of sizes $1040$, $2532$ and $4560$, the deficit certificates, the radius-six integer certificate, the 96-point table, the doubled-zonotope shells for $n\le4$, the coordinate-segment identity at $46$ parameter pairs, and the three-dimensional zonotope with its 392-point forced set. It uses only Python's standard library and runs in a few seconds. The ring counts for $7\le n\le10$ and the enumeration of Section~\ref{sec:open} were obtained with the NumPy scripts of the computational record.

\section{The coordinate-segment calculation}\label{app:segments}
We prove Proposition~\ref{prop:segment}. Normalize $H_n\subseteq[0,W]^2$. Its bottom edge is $[a,a+n]\times\{0\}$, where $a=(W-n)/2$. The adjacent primitive directions are $(n,1)$ and $(-n,1)$; all boundary points off the horizontal edges belong to the left chain $x\le a-n$ or the right chain $x\ge a+2n$. These descriptions include the corresponding upper chains by reflection.

In $P_t=H_n+[0,t]e_1$, each nonhorizontal edge retains its length and all its lattice points belong to $Z_n(P_t)$. On a horizontal edge, now of length $n+t$, the forced boundary centres are the points at most $n$ steps from either endpoint.

Let $U_t$ be the union in \eqref{eq:tube}. In any row $0\le y\le n$, the left-chain balls and the selected left end of the bottom edge cover from the left end of the hull row through $a+2n$. The right-chain balls and selected right end cover from $a+t-n$ through the right end. There are no internal gaps in either covered portion, since consecutive boundary centres differ by at most $n$ in sup norm. The middle has $\max(0,t-3n-1)$ uncovered lattice points, while the hull row has gained exactly $t$ points. Hence the union grows by $\min(t,3n+1)$ in each of these $n+1$ rows, and likewise in the $n+1$ top rows. The two groups are disjoint because $W\ge3n$.

In the remaining rows, horizontal-edge balls do not contribute. The left and right covered portions are disjoint already at $t=0$ and are translated apart as $t$ grows, so their counts are unchanged. Therefore
\[
|U_t|=A_n+2(n+1)\min(t,3n+1).
\]
All nonhorizontal selected centres have zero slack. On each horizontal edge, the selected positions are the distinct integers in $[0,n]\cup[t,n+t]$, and their slack sum is
\[
\sum_j\bigl(\min(j,n)+\min(n+t-j,n)-n\bigr)
=n\min(t,n+1).
\]
Subtracting the two horizontal-edge sums gives the stated formula. For $1\le t\le n+1$ the increase is exactly $2t$; thereafter it increases and finally stabilizes at $2(n+1)(2n+1)$.

\end{document}